\documentclass[a4paper, fleqn,usenatbib]{mnras}

\usepackage{mathptmx}

\usepackage[T1]{fontenc}
\usepackage{ae,aecompl}

\def\be{\begin{equation}}
\def\ee{\end{equation}}

\usepackage{graphicx}	
\usepackage{amsmath}	
\usepackage{amssymb}	

\title[Pulsar white dwarfs]{AR Scorpii and possible gravitational wave radiation from pulsar white dwarfs}

\author[B. Franzon et al.]{
B. Franzon, $^{1}$\thanks{E-mail: franzon@fias.uni-frankfurt.de}
S. Schramm,$^{2}$\thanks{E-mail: schramm@fias.uni-frankfurt.de}
\\
$^{1,2}$Frankfurt Institute for Advanced Studies, Ruth-Moufang-1 60438 Frankfurt am Main, Germany\\
}

\date{Accepted XXX. Received YYY; in original form ZZZ}

\pubyear{2016}

\begin{document}
\label{firstpage}
\pagerange{\pageref{firstpage}--\pageref{lastpage}}
\maketitle

\begin{abstract}
In view of the new recent observation and measurement of the rotating and highly-magnetized white dwarf AR Scorpii \cite{Marsh:2016uhc}, we determine bounds of its moment of inertia, magnetic fields and radius. Moreover, we investigate the possibility of fast rotating and/or magnetized white dwarfs to be source of detectable gravitational wave (GW) emission. Numerical stellar models at different baryon masses are constructed.  For each star configuration, we compute self-consistent relativistic solutions for white dwarfs endowed with poloidal magnetic fields by solving the Einstein-Maxwell field equations in a self-consistent way. The magnetic field supplies an anisotropic pressure, leading to the braking of the spherical symmetry of the star. In this case, we compute the quadrupole moment of the mass distribution. Next, we perform an estimate of the GW of such objects. Finally, we show that the new recent observation and measurement pulsar white dwarf AR Scorpii, as well as other stellar models, might generate gravitational wave radiation that lies in the bandwidth of the discussed next generation of space-based GW detectors DECI-hertz interferometer Gravitational wave Observatory (DECIGO) and Big Bang Observer (BBO).

\end{abstract}


%

\section{Introduction}
Some white dwarfs (WD) are associated with strong magnetic fields. From observations, it was shown that the surface magnetic field in these stars can reach values up to $10^{9}\,$ G, see  \cite{Terada:2007br,Reimers:1995ia,Schmidt:1995eh,Kemp:1970zz,putney1995three,angel1978magnetic}. However, the internal magnetic field in stars is very poorly constrained by observations and can be much stronger than the one at the surface. A virial-theorem based estimate by equating the magnetic field energy with the gravitational biding energy leads to an upper limit for the magnetic fields inside WD's of  $\sim 10^{13}$ G  \citep{shapiro2008black}. On the other hand, in the context of exploring overluminous type Ia supernovae and the possible existence of super-Chandrasekhar white dwarfs, different numerical calculations suggest that white dwarfs might have internal magnetic fields as large as $10^{12-16}\,$ G \citep{das2013new, Bera:2015yxa, Franzon:2015gda}. 

Due to their large radius compared to neutron stars, white dwarfs present a different scale in their macroscopic stellar properties, like the moment of inertia, magnetic moment, rotational frequency and quadruple moment. This, together with the fact that white dwarfs are found closer to earth, is one of the reasons why white dwarfs are much more understood than neutron stars.

The main motivation of this work is provided by the recent binary system  discovered by \cite{Marsh:2016uhc}. This system is composed of a main sequence star and a fast spinning and magnetized white dwarf with a mass that lies in the range $0.81M_{\odot}<M_{\rm{wd}}\leq1.29M_{\odot}$. The pulsating white dwarf is called AR Scorpii, AR Sco for short, and such a system has never been observed orbiting a cool red dwarf star. In the case of AR Sco, this was the first radio pulsations detected in any white dwarf system. In addition, due to its high magnetic field combined with rotation, such type of WD's  are called  `white dwarf pulsar'.

The rapidly-spinning and magnetized stellar remnant  AR Sco pulses across almost the entire electromagnetic spectrum, from X-ray to radio wavelengths. Up to now, pulsating stars were related to neutron stars (NS), which can be rotating and  highly magnetized, emitting a beam of electromagnetic radiation. Typically,  white dwarfs rotate with periods of days or even years. On the other hand, according to \cite{Mereghetti:2010id}, one of the fastest observed WD possesses a spin period of $13.2\,s$, a value similar to the ones observed in Soft Gamma Repeaters (SGR) and  Anomalous X-ray pulsars (AXP), known as magnetars \citep{Duncan:1992hi, Thompson:1993hn}. A relation between WD's and magnetars was addressed  by \cite{Malheiro:2015yda}, where the authors speculated that SGR's and AXP's with low surface magnetic field might be rotating magnetized white dwarfs. 

As in neutron stars, the origin of magnetic fields in white dwarfs is still under debate. While the magnetic flux conservation of a progenitor remains an attractive possibility, a likely origin of the such strong magnetic fields is a dynamo process that operates during the envelope evolution, see \cite{ferrario2016magnetic}. Recently, observations have shown that  the formation  of  high  magnetic  white  dwarfs  can be  related to  strong  binary  interactions  during  post-main-sequence  phases  of stellar evolution \citep{nordhaus2011formation}. 

Whatever the origin of strong magnetic fields might be, they effect the stars in different ways. First, magnetic fields affect locally the microphysics of the equation of state through the Landau quantization of the energy levels of charged particles.  However, as already shown by \cite{bera2014mass}, although equation of state of electron degenerate matter is strongly modified due to Landau quantization, this effect is negligible on the global properties of white dwarfs. In fact, we already shown that in NS's the  contribution to the structure of the star when taking into account the magnetic field corrections in the equation of state is very small, see \cite{franzon2016self}. For this reason, we do not take into consideration a magnetic-field-dependent equation of state in our calculation.

Secondly, magnetic fields are sources of the gravitational field equations through the Maxwell energy-momentum tensor. As a result, they make the pressure anisotropic, which requires a general treatment beyond the Tolman-Oppenheimer-Volkoff (TOV) solutions \citep{oppenheimer1939massive, tolman1939static}. In addition, the Lorentz force induced by magnetic fields change the structure of white dwarfs, which, in the case of poloidal fields, become oblate objects. This is the same effect as the one produced by rotation. In both cases, the star deformation with respect to the magnetic and/or rotation axis can be quantified by the stellar quadrupole moment.   

As predicted by \cite{einstein1916approximative}, gravitational waves are generated by objects that have quadrupole moment varying in time, such as colliding black holes, collapse of stellar cores, coalescing neutron stars, white dwarf stars, etc. Such systems disrupt the space-time producing  GW that radiate from the source and travel at the speed of light through the Universe, carrying  information about their sources, as well as the nature of gravity itself.

Currently, the main ground-based gravitational waves interferometer  operating is  the twin Laser Interferometer Gravitational-wave Observatory (LIGO) which sensitivity is designed to detect GW amplitude of one part in $10^{21}$ within the frequency bandwidth in the range 30 - 7000 Hz.  In the next years, a second generation of detectors, as for example, 
advanced-LIGO and advanced-Virgo, will be operating.   
Furthermore, the space-based gravitational waves detector Laser Interferometer Space Antenna (LISA) \citep{danzmann1996lisa} has been planning to be launched. LISA operates a space-based gravitational waves detector sensitive at frequencies between 0.03 mHz and 0.1 Hz. 

The Deci-Hertz Interferometer Gravitational Wave Observatory (DECIGO) \citep{seto2001possibility, kawamura2006japanese} is a plan of a future Japanese space mission for observing GW's in frequency bandwidth similar to LISA, however, at lower gravitational waves amplitudes. This fact, as we are going to see, makes DEGICO suitable to detect gravitational waves from fast rotating and/or magnetized white dwarfs. Meanwhile, another space-based interferometer has been proposed as a successor to LISA, the Big Bang Observer (BBO) \citep{phinney2003big}, with both frequency bandwidth and gravitational waves amplitudes similar to the ones of DECIGO. 

With this in mind, we make use of available data of AR Sco, as its distance from Earth, its rotation frequency and its mass range, to perform self-consistent rotating and magnetized white dwarf calculations and then determine bounds of its radius, moment of inertia, quadruple moment and magnetic fields. With these results, we investigate the possibility of rotating and magnetized white dwarfs to be sources of detectable GW emission. 

A white dwarf is a dense configuration supported by the electron degeneracy pressure against gravitational collapse. Here, we describe the stellar interior assuming that the star is predominately composed of carbon $^{12}C$ ($A/Z = 2$) in an electron background, see \cite{chandrasekhar1931maximum}. In fact, white dwarfs are much less compact stars than neutron stars, being easily deformed due to magnetic fields. Another important point is that  white dwarfs can be considered the most closest astrophysical sources of gravitational waves.

We organise the paper in the following way.In Sec. 2 we summarise the coupled Maxwell-Einstein equations and the hydrodynamic equations in presence of a magnetic field in general relativity. In Sec. 3, we describe the results concerning the influence of strong magnetic fields on the structure of white dwarfs, in particular to the AR Sco star. Sec. 4 describes the effects of rotation, combined also with magnetic fields, on the detectability of gravitational waves
emitted by rotating and/or magnetized white dwarfs. Finally in Sec. 5, we give our conclusions. 

\section{White dwarfs with axisymmetric magnetic fields} 
We follow the numerical technique as developed by \cite{Bonazzola:1993zz, Bocquet:1995je} to obtain magnetized white dwarf configurations in a fully  general relativity way. Within this approach, the Einstein-Maxwell equations are solved numerically by means of a pseudo-spectra method for axisymmetric stellar configurations within the 3+1 formalism in General Relativity. The accuracy of solutions is controlled by a 2D general-relativistic virial theorem \citep{Bonazzola:1993zz} and, for the stars presented here, it is typically of the order of $10^{-5}$. Recently, this formalism was applied  to neutron stars by \cite{franzon2016self, Franzon:2016iai, Franzon:2016urz} and to study magnetized white dwarfs by \cite{Franzon:2015gda}. 

With the assumption of a stationary and axisymmetric spacetime, the line element in spherical-like coordinates $(r, \theta, \phi)$ can be written as:
\begin{align}
ds^{2} = &-N^{2} dt^{2} + \Psi^{2} r^{2} \sin^{2}\theta (d\phi - \omega dt)^{2} \\
 &+ \lambda^{2}(dr^{2} + r^{2}d\theta^{2}), \nonumber
\label{line}
\end{align}
with N, $\omega$, $\Psi$ and $\lambda$ being functions of $(r, \theta)$. 
The gravitational field is deduced from the integration of a coupled system of four elliptic partial differential equations for the four metric functions, see e.g. \cite{Bonazzola:1993zz}. The final system of gravitational equations can be cast in the form:

\be 
\Delta_{2} [(N \Psi-1) r \sin \theta] = 8\pi N\lambda^2 \Psi r \sin \theta (S^{r}_{r} + S^{\theta}_{\theta}),
\label{Bfinal}
\ee

\be 
\Delta_{2} [{\rm{ln}} \lambda + \nu] = 8\pi \lambda^2 S^{\phi}_{\phi} + \frac{3 \Psi^2 r^2 \sin^2 \theta}{4 N^2} \partial \omega \partial \omega - \partial \nu \partial \nu,
\label{Afinal}
\ee

\be 
\Delta_{3} \nu = 4\pi \lambda^2 (E + S) + \frac{\Psi^2 r^2 \sin^2 \theta}{2N^2} \partial \omega \partial \omega - \partial \nu \partial( \nu + {\rm{ln}} \Psi),
\label{Nfinal}
\ee

and 

\begin{align}
&\left[ \Delta_{3} - \frac{1}{r^2 \sin^2 \theta} \right] (\omega r \sin \theta) \nonumber \\ = & -16\pi \frac{N\lambda^2}{\Psi^2} \frac{J_{\phi}}{r \sin \theta}+ r \sin \theta  \partial \omega \partial(\nu - 3 {\rm{ln}} \Psi),
\label{omegafinal}
\end{align}
where the short notation was introduced:

\begin{align}
& \Delta_{2} = \frac{ \partial^2}{\partial r^2} + \frac{1}{r}\frac{ \partial}{\partial r} + \frac{1}{r^2}\frac{ \partial^2}{\partial \theta^2} \nonumber\\
& \Delta_{3} = \frac{ \partial^2}{\partial r^2} + \frac{2}{r}\frac{ \partial}{\partial r} + \frac{1}{r^2}\frac{ \partial^2}{\partial \theta^2} + \frac{1}{r^2 \tan \theta}\frac{ \partial}{\partial \theta} \nonumber \\
& \nu = {\rm{ln}} N \nonumber.
\end{align}

In addition, in the final gravitational field equations system, Eqs.~\ref{Bfinal}-\ref{omegafinal}, terms as $\partial \omega \partial \omega$ are defined as: 

\be 
\partial \omega \partial \omega := \frac{\partial \omega}{\partial r}\frac{\partial \omega}{\partial r} + \frac{1}{r^2}\frac{\partial \omega}{\partial \theta}\frac{\partial \omega}{\partial \theta} \nonumber, 
\label{notation}
\ee
and the total energy density, momentum density and stress tensors of the system are:
\be 
E = \Gamma^{2} \left( e + p \right) - p + E^{EM}
\ee

\be 
J_{\phi} = \left( E + p \right) \lambda^{2} \Psi  r \sin \theta U + J_{\phi}^{EM}
\ee

\be 
S^{r}_{r} = p + S^{EM\,\,r}_{r} 
\ee 
\be 
 S^{\theta}_{\theta} = p +  S^{EM\,\,\theta}_{\theta}, 
\ee 

\be 
S^{\phi}_{\phi} = p + \left( E + p \right) U^2 + S^{EM\,\,\phi}_{\phi},
\ee
with $e$ and $p$ being the energy and the isotropic pressure of the fluid, $U$ the fluid velocity, $\Gamma$  represents the Lorentz factor which relates the Eulerian and the fluid comoving observers, and $E^{EM}$, $J_{\phi}^{EM}$,  $S^{EM\,\,r}_{r}$, $ S^{EM\,\,\theta}_{\theta}$ and $S^{EM\,\,\phi}_{\phi}$ correspond to the electromagnetic contribution to the energy, momentum and stress tensor of the system, see e.g. \cite{franzon2016self,Bocquet:1995je}.

According to \cite{salgado1994high}, global quantities as the gravitational mass $M$ and the quadrupole momentum $Q$ are identified as the leading term in the asymptotic behavior of the metric potential $N(r, \theta)$:

\be 
{\rm{ln}}\, N(r, \theta)_{r\rightarrow \infty} \sim -\frac{M}{r} + \frac{Q}{r^3} P_{2} (cos \theta),
\label{}
\ee
with $P_{2} (cos \theta)$ being the second order Legendre polynomial. On the other hand, according to \cite{Bonazzola:1993zz, salgado1994high}, the quadruple moment can be written as:

\be 
Q = -\frac{1}{4\pi} \int \sigma_{{\rm{ln}}N} P_{2} (cos \theta) r^4 \sin \theta dr d\theta d\phi,
\label{quadru}
\ee
where $\sigma_{{\rm{ln}}N}$ is the source term in Eq.~\ref{Nfinal} (see also Eq. (3.19) in \cite{Bonazzola:1993zz}).

%

In the case of rigid rotation, the equation of motion ($\nabla_{\mu} T^{\mu\nu}=0$) of a star endowed with magnetic fields
yields:
\be
H \left(r, \theta \right) + \nu \left(r, \theta \right)  + M \left(r, \theta \right) - \Gamma (r, \theta) = const,
\label{equationofmotion}
\ee
where $H(r,\theta)$  is the heat function defined in terms of the baryon number density $n$:
\be
H = \int^{n}_{0}\frac{1}{e(n_{1})+p(n_{1})}\frac{d P}{dn}(n_{1})dn_{1},
\label{heat}
\ee
and the magnetic potential $M(r,\theta)$ can be expressed as:
\be
M \left(r, \theta \right) = M \left( A_{\phi} \left(r, \theta \right) \right): = - \int^{0}_{A_{\phi}\left(r, \theta \right)} f\left(x\right) \mathrm{d}x,
\ee
with $A_{\phi}$ being the magnetic vector potential and $f(x)$ an arbitrary function $f(x)$ that needs to be chosen \citep{Bocquet:1995je}. In our case, we construct stellar models for constant values of $f(x)=f_{0}$. The macroscopic electric current relates to $f_{0}$ as $j^{\phi} \propto (e + p)\,f_{0}$. Therefore, for higher values of the current function, the electric current increases and, therefore, the magnetic field in the star increases proportionally.  As already shown  by \cite{Bocquet:1995je}, the current function can have a more complex structure, however the general conclusions remain the same.   

\section{Global properties of pulsar white dwarf AR Scorpii}
Stationary and axisymmetric configurations are calculated once the EoS is specified, together with the rotation law, the central energy density and the current function $f_{0}$. In our case, we chose to construct uniformly rotating and/or magnetized models at a given baryon mass by varying the star frequency and also $f_{0}$. In our first analysis, we chose to study stellar configurations with masses between the maximum and the minimum possible stellar masses for the white dwarf AR Sco, i.e., $M_{B}=1.29, 1.18, 1.00, 0.81 M_{\odot}$. Furthermore, we fix the stellar frequency at $\nu=8.538$ mHz, which is the observed frequency of this star. In fact, our results are general and the approach shown here can be also applied to other systems.  

\begin{figure}
\begin{center}
\includegraphics[width=0.68\textwidth,angle=-90,scale=0.5]   {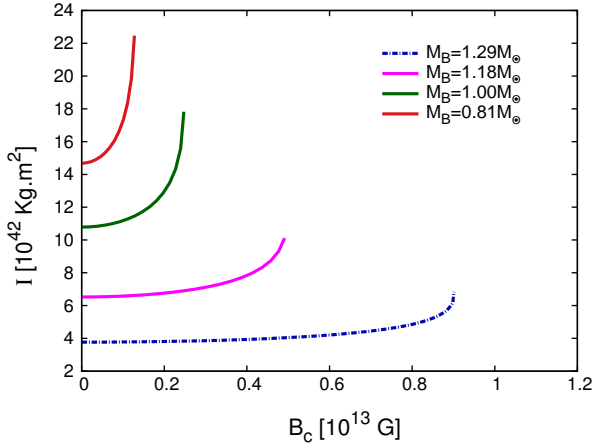} \quad
\caption{Moment of inertia as a function of central magnetic fields for four stellar configurations at different fixed baryon masses. These stars rotate at the frequency of $\nu=8.538$ mHz.} 
\label{bc_iner}
\end{center}
\end{figure}

In Fig.~\ref{bc_iner}, we show the moment of inertia, $I$, as a function of central magnetic fields, $B_{c}$, for stellar configurations at different fixed baryon masses. The same behavior as shown in Fig.~\ref{bc_iner} was obtained for $I$ as a function of polar surface fields, $B_{s}$. $I$ increases because magnetic fields deform the star, which becomes larger in the equatorial plane. In Fig.~\ref{bc_iner}, the purely rotating contribution to the moment of inertia of the system is found when $B_{c}=0$. On this same figure, the end points on the curves correspond to maximum magnetic field configurations achieved within the code. At this point, the star becomes highly magnetized and deformed, reaching a doughnut-shaped density distribution, see \cite{cardall2001effects}. However, the numeric techniques described in section II does not handle toroidal-shape solutions, which gives us a practical limit within this approach. 

According to Fig.~\ref{bc_iner}, heavier WD's can support higher magnetic fields in their interiors. This is also seen from observation, where the majority of observed WD's are more massive than non-magnetized ones, see e.g. \cite{Boshkayev:2014kua} and references therein. This is simply because massive stars are more dense in their interiors, and in ideal magneto hydrodynamics (MHD), the magnetic field is 'frozen' with the fluid and, therefore, it is proportional to the local mass density of the fluid  \citep{mestel2012stellar}. Still according to Fig.~\ref{bc_iner},  heavier WD's have smaller moment of inertia. This is due to the fact the big radii of lighter white dwarfs significantly contribute to the momentum of inertia of stars. For example,
a star with $M_{B}=1.29\,M_{\odot}$ has a maximum moment of inertia of $I=6.83\times 10^{42} \,\,{\rm{kg.m^2}}$. In this case, the central magnetic field is $B_{c}=0.90\times 10^{13}$ G, with the corresponding surface magnetic field at the pole of $B_{s}=8.95\times 10^{11}$ G. On the other hand, a less massive star at a fixed baryon of $M_{B}=0.81\,M_{\odot}$ has a maximum moment of inertia of $I= 22.47\times 10^{42} \,\,{\rm{kg.m^2}}$, which is found when the central magnetic field is $B_{c}=0.13\times 10^{13}$ G and $B_{s}=2.14\times 10^{11}$ G on the surface. 

\begin{figure}
\includegraphics[width=0.68\textwidth,angle=-90,scale=0.5]{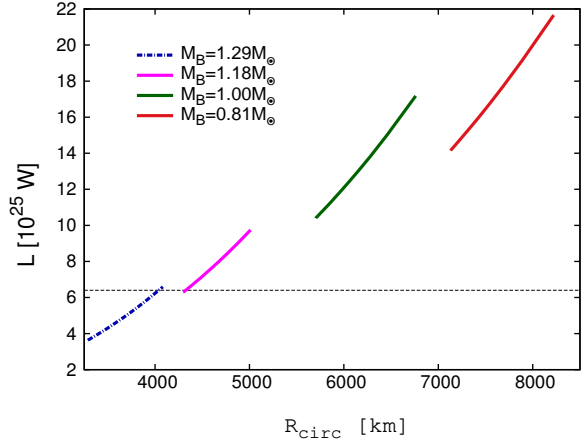}
\caption{Spin down luminosity $L$ as a function of the circular equatorial radius $R_{circ}$ for the same configurations as shown in Fig.~\ref{bc_iner}. The horizontal dashed line represents the maximum luminosity of the white dwarf AR Sco.}
\label{rc_luminosity}
\end{figure}

In Fig.~\ref{rc_luminosity} we show the  spin-down luminosity $L$ as a function of the circular equatorial radius for the same stars as already depicted in Fig.~\ref{bc_iner}. In this case, the spinning star loses energy at a rate $L=-4\pi^2\nu \dot{\nu} I$ \citep{Marsh:2016uhc}, with $I$ being the moment of inertia, which is now a function of the magnetic field $I(B)$. \cite{Marsh:2016uhc} reported also a stellar maximum luminosity for AR Sco to be about $L_{obs}\sim 6.3\times10^{25}$ W, which it is represented by the dashed and black horizontal line in Fig.~\ref{rc_luminosity}. 

For a star with $M_{B}=1.29\,M_{\odot}$, and using the AR Sco observed values for the frequency $\nu=8.538$ mHz and the frequency derivative $\dot{\nu}=-2.86\times10^{-17}$ Hz $s^{-1}$ \citep{Marsh:2016uhc}, we obtain a stellar maximum luminosity of $L \sim 6.4\times10^{25}$ W, whose value  matches the experimental maximum luminosity of AR Sco (horizontal dashed line in Fig.~\ref{rc_luminosity}) for a star with $M_{B}=1.29\,M_{\odot}$. Moreover, the corresponding radius of this star is $R_{circ}=4080.00$ km,  for a central magnetic field of $B_{c}\sim 0.90\times 10^{13}$ G, which is almost the maximum magnetic field reached at the center of a star with $M_{B}=1.29\,M_{\odot}$, see Fig.~\ref{bc_iner}.

Nevertheless, any point on a curve lying above the horizontal line of Fig.~\ref{rc_luminosity} could in principle explain the observed luminosity of AR Sco, and thus any mass lower than $M_{B}=1.29\,M_{\odot}$, with even a very low magnetic field (or even for non-magnetized white dwarf). For example, the star with $M_{B}=1.18\,M_{\odot}$ has a corresponding spin-down luminosity of $L\sim L_{obs}$ for  central and surface magnetic fields of $B_{c}= 1.25\times 10^{11}\,$ G and $B_{s}= 9\times 10^{9}\,$ G, respectively. This value of $B_{s}$, as well surface fields obtained with other choices of (lower) stellar masses, are compatible to values of surface magnetic fields experimentally observed in white dwarfs. Note that, a lower magnetic field implies a much lower radius, moment of inertia and, as we will see, a lower gravitational wave amplitude. According to Fig. 2, if the white dwarf has a higher observed luminosity, lower masses stars would be favored to explain the AR Sco data.  

Magnetic fields on the stellar surface can be estimated from observations of the star's period and period derivative. In this case, one considers a magnetic dipole which rotates and  emits  electromagnetic radiation. As a consequence, the star loses energy and spin down. However, one  usually uses fidudial parameters that do not depend on the magnetic field  (for example,  a constant moment of inertia $I$) to estimate polar surface magnetic fields through the magnetic dipole formula. For a star at $M_{B}=1.18\,M_{\odot}$, a trivial estimate of the magnetic field assuming a magneto-dipole radiation losses gives for the surface magnetic field $B_{s}\sim 10^{9}\,$ G.  In this work, as we take into consideration the effect of magnetic fields on global properties of stars, we obtain surface magnetic fields higher than just the dipole model. 

\section{Gravitational waves from white dwarfs }

The Lorentz and the centrifugal forces induced by magnetic fields and rotation make the star oblate. In this case, the stellar deformation can be quantified by the stellar quadrupole moment, $Q$, measured with respect to the rotational axis, for example. According to \cite{bonazzola1996gravitational}, the gravitational wave amplitude $h_{0}$ of a perpendicular rotator can be estimated as:
\begin{equation}
h_{0}= \frac{6G}{c^{4}}\frac{\Omega^{2}}{d}Q,
\label{gw}
\end{equation}
with $G$ being the gravitational constant, $c$ the speed of light,  $d$ the distance of the star, and $\Omega$ the rotation velocity of the star. For all stars considered in this work, we use a distance $d\sim$116 pc (1pc = 3.08$\times 10^{13}\,$km) which is the distance of AR Sco from Earth, see \cite{Marsh:2016uhc}. Although we are fixing $d$, note that the closer the star, the larger the amplitude $h_{0}$.

By means of the rotating and magnetic white dwarfs calculations as already shown in Fig.~\ref{bc_iner}, we make a crude estimate of the GW strength emitted by these objects. This is crude because the white dwarfs as computed here have rotation and magnetic axes aligned. In this case, even stars strongly deformed do not emit gravitational radiation  \citep{bonazzola1996gravitational}. However, an estimate of the strength of gravitational wave emission can be deduced if we assume that the magnetic axis and the rotation axis are not aligned. This is a good approximation since the rotational and magnetic field effects do not compete in deforming the star because of the rather slow rotation.

\begin{figure}
\vspace{0.1cm}\includegraphics[width=0.68\textwidth,angle=-90,scale=0.5]{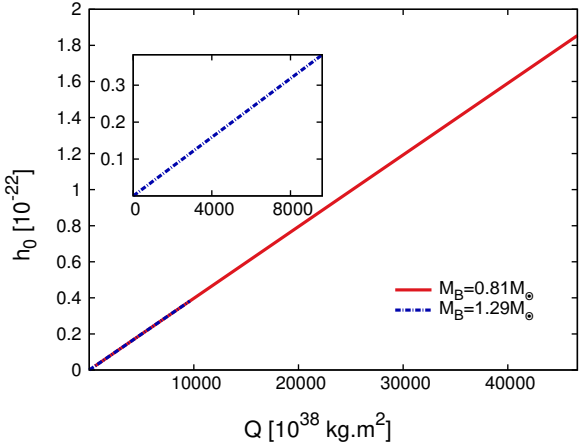}
\caption{Gravitational wave amplitude $h_{0}$ as a function of the quadrupole moment $Q$ for two stars at fixed baryon masses of $M_{B}=0.81\,M_{\odot}$ and $M_{B}=1.29\,M_{\odot}$, respectively. Both star rotate at a frequency of $\nu=8.538$ mHz.}
\label{h0_quadr}
\end{figure}

Fig.~\ref{h0_quadr} depicts the gravitational wave amplitude $h_{0}$ for white dwarfs at $M_{B}=0.81\,M_{\odot}$ and $M_{B}=1.29\,M_{\odot}$, respectively. The higher the stellar mass, the smaller the $h_{0}$. This is due to the fact that a star with $M_{B}=0.81\,M_{\odot}$ has lower density that is spread out over a larger radius. As a result, the star easily becomes deformed through magnetic field effects and, therefore, its quadrupole moment is higher than in more massive white dwarfs. For example, a white dwarf at $M_{B}=0.81\,M_{\odot}$ possesses a maximum GW amplitude of $h_{0}=1.85\times10^{-22}$, while
we obtain a maximum value of $h_{0}=3.82\times 10^{-23}$ for the star $M_{B}=1.29\,M_{\odot}$. Although in a different frequency band, these values are only one and two order of magnitude lower than the first direct
detection of gravitational waves from black hole merger, see \cite{abbott2016observation}. Note that, white dwarfs with lower masses have gravitational wave amplitude of about three order of magnitude higher than slow rotating and highly magnetized neutron stars, see \cite{Franzon:2016urz, bonazzola1996gravitational, bonazzola1994astrophysical}. This is related to the scale of the quadrupole moment, which is of the order of $Q\sim10^{42}\,{\rm{kg.m^2}}$ in magnetic white dwarfs, while this value reduces to $\sim10^{38}\,{\rm{kg.m^2}}$ for neutron stars. In addition, the WD's considered here are much closer to Earth than a typical neutron star, contributing to make them good sources for gravitational wave emission and detection. 

\begin{figure}
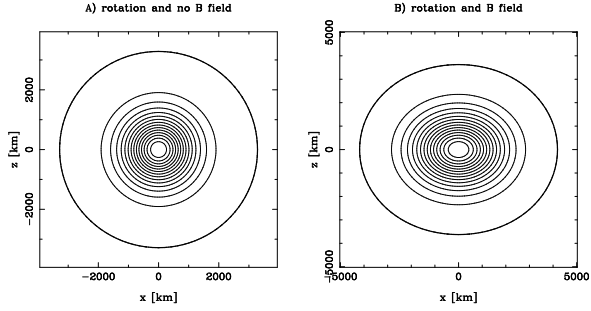

\begin{center}
\includegraphics[width=0.45\textwidth,angle=-90,scale=0.5]{density_rot_noB.eps} \quad
\includegraphics[width=0.45\textwidth,angle=-90,scale=0.5]{density_rot_B.eps} 
\caption{Baryon number density profile  in the meridional plane (x, z) of a star with $M_{B}=1.29\,M_{\odot}$ A) rotating at a frequency of 8.538 mHz and without magnetic fields (left panel), and B) highly magnetized and rotating at the same frequency as in A) (right panel). In the first case, the ratio between the polar and equatorial radius is $r_{p}/r_{eq}=0.99$, while for the magnetized white dwarf we find $r_{p}/r_{eq}=0.84$. The magnetic dipole moment in case B) reaches 2.53$\times 10^{34}$ $\rm{Am^{2}}$. In our case, the magnetic field  and the rotation axis are aligned (z direction).}
\label{density}
\end{center}
\end{figure}

Fig.~\ref{density} shows the baryon number density profile for a star at $M_{B}=1.29\,M_{\odot}$, but in two different scenarios: A) rotating at a frequency of $\nu=8.538$ mHz, but without magnetic fields and B) rotating at the same frequency as in A), but endowed with magnetic fields. In the case A), the deviation from spherical symmetry is very small, with the polar radius being 0.05$\%$ smaller than the equatorial radius. On the other hand, in the case B), the circular equatorial radius is 4080.00 km in contrast to 3288.10 km in case A). The gravitational wave amplitude in case A) is $h_{0}=2.46\times10^{-26}$, while this value raises to $h_{0}=3.82\times10^{-23}$ with magnetic fields.  In face of this, one sees that the effects of magnetic fields in deforming the star and, therefore, contributing strongly to GW emission is evident. Furthermore, one sees that the deviations from spherical symmetry are remarkable and need to be taken into consideration while modeling highly magnetized white dwarfs. As a consequence, a simple TOV solution can not be applied in these cases. 
  
As it was already discussed, strong magnetic fields do increase the quadruple moment of stars. For example, a rotating and non-magnetized white dwarf, case A in Fig.~\ref{density}, has  a quadruple moment of $Q=6.20\times 10^{38}\,{\rm{kg.m^2}}$, while its magnetized counterpart, case B) in Fig.~\ref{density}, supports a quadrupole moment of $Q=1.12\times 10^{42}\,{\rm{kg.m^2}}$. Again, this is due to the Lorentz force associated with the macroscopic currents that generate
the field, which pushes the matter off-center. As a result, the  stellar equatorial radius increases  and the star becomes much more deformed with respect to the symmetry axis. In this case, the baryon number density, $n_{b}$, reached at the center of the star is $6.30\times 10^{7}\,{\rm{g/cm^3}}$, which is one order of magnitude smaller than the baryon number density of a purely rotating  white dwarf, case A), which reaches $n_{b}=3.12\times 10^{8}\,{\rm{g/cm^3}}$ at its center.
 
\begin{figure}
\vspace{0.1cm}\includegraphics[width=0.68\textwidth,angle=-90,scale=0.5]{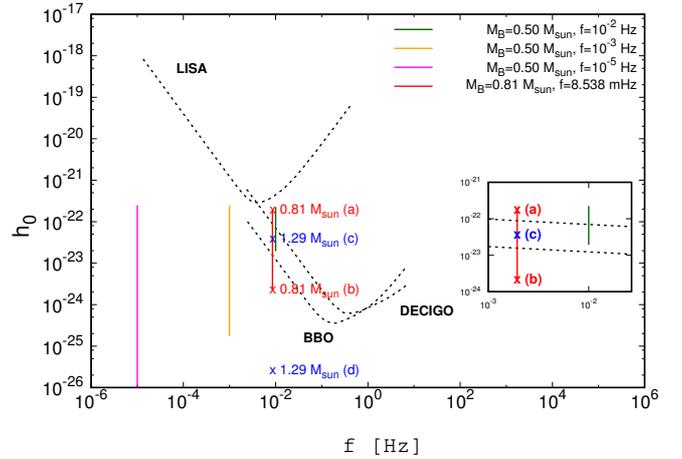}
\caption{The solid black curves represent the spectra for LISA, DECIGO and BBO interferometers. The vertical lines are rotating white dwarfs at fixed baryon masses. The models (a), (b), (c) and (d) are estimates of gravitational wave amplitudes by using the supposed masses for the pulsar white dwarf AR Scorpii.}
\label{plot}
\end{figure}

In Fig.~\ref{plot}, we show the frequency bandwidth of different space-borne gravitational wave interferometers, the Laser Interferometer Space Antenna (LISA), BBO and DECIGO. On this figure, we show also the estimate of gravitational wave amplitude for the same stars as shown in Fig.~\ref{h0_quadr}. First, a white dwarf at $M_{B}=1.29\,M_{\odot}$, purely rotating at 8.538 mHz, is represented by the letter (d), and has $h_{0}$ lower than those ones expected to be detected by BBO and DECIGO. This star, however, has its gravitational wave amplitude raised with the inclusion of magnetic fields, case (c), and can, potentially, be detected by BBO. This behavior is similar to the case of a white dwarf at  $M_{B}=0.81\,M_{\odot}$ (red line), which lie in the amplitude range of both BBO and DECIGO when magnetized, case (a), not being, however, a potential candidate when purely rotating, case (b). 

In order to investigate stellar models different from AR Sco, as well if purely rotating white dwarfs are able to produce detectable GW, we include in Fig.~\ref{plot} calculations for magnetized and rotating white dwarf models at fixed baryon mass of $M_{B}=0.50\,M_{\odot}$. This star rotates at different rotation frequencies of $f=10^{-5}$ Hz (magenta line), $10^{-3}$ Hz (orange line) and $10^{-2}$ Hz (green line), respectively. It is interesting to see that the higher the frequency, the smaller the range of the GW amplitude $h_{0}$. This is due to the fact that the rotational frequency limits further the density of the star and, therefore, the magnetic field. 


As can be seen from Fig.~\ref{plot}, the white dwarf $M_{B}=0.50\,M_{\odot}$ rotating at $f=10^{-2}$ Hz, and without magnetic fields (green line), can clearly be in the range of detectability of the BBO detector. In this case, the rotating white dwarf has a GW amplitude of $\sim 7 \times 10^{-23}$. In addition, if this star is magnetized,  this would lead to a maximum GW amplitude of the order of $2\times 10^{-22}$, which is in the range of detectability of both BBO and DECIGO interferometers.

\begin{figure}
 \vspace{0.1cm}\includegraphics[width=0.68\textwidth,angle=-90,scale=0.5]{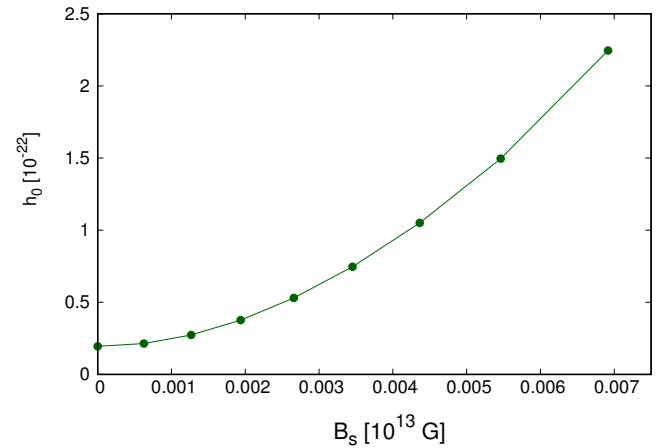}
\caption{Gravitational wave amplitude as a function of the surface magnetic field for a white dwarf at fixed baryon mass of $M_{B}=0.50\,M_{\odot}$ rotating at a frequency of $10^{-2}$ Hz.}
\label{hbs}
\end{figure}

In Fig.~\ref{hbs}, we show the GW amplitude $h_{0}$ as a function of surface magnetic fields, $B_{s}$, for the  star $M_{B}=0.50\,M_{\odot}$  (green line) as depicted in Fig.~\ref{plot}. We chose to show $B_{s}$  instead the magnetic dipole moment or the central magnetic field since this is the quantity that can be observed in white dwarfs. Note that, according to Fig.~\ref{plot} and Fig.~\ref{hbs}, BBO is also able to detect GW from stars with low magnetic fields. The results of this analysis can be generalized to other frequencies or masses.  However, as the frequency increases, the star approaches to the  mass-shedding (Kepler) limit, which for the white dwarf $M_{B}=0.50\,M_{\odot}$ is $f\sim 0.02$ Hz. 

It is instructive to compare the white dwarf luminosity due to emission of electromagnetic waves (assuming vacuum dipole emission) and the luminosity due to gravitation waves emission. The luminosity due to emission of electromagnetic waves of a perpendicular rotator can be written as $L^{dip} = (2/3) R^6 B^2 \Omega^4$, see e.g. \cite{shapiro2008black}. By using values similar to those as depicted in Fig.~\ref{hbs}, we have for the spin-down luminosity  $L\sim 10^{21}$ W and $ L^{dip}\sim  10^{21}$ W. On the other hand, by using the luminosity due to gravitational waves of a given source as shown in  \cite{kokkotas2002gravitational}, we obtain $L^{GW}\sim 10^{19}$ W. Therefore, the spin-down luminosity of the star is mainly due to dipole radiation.
  
According to Fig.~\ref{hbs}, the magnetic field that corresponds to the minimum GW amplitude, $h_{0}\sim 2\times 10^{-22}$, that lies in the DECICO range of detectability, is $\sim 3.2\times 10^{10}$ G. More importantly, this value is just one order of magnitude smaller than surface magnetic field already observed in white dwarfs. Therefore, both rotating and magnetized white dwarfs could be good candidates of future GW observation and detection by the new generation of GW interferometers. As already discussed, the gravitational wave amplitude $h_0$ can be obtained in the regime of slow rotation and strong magnetic fields from Eq.~\ref{gw}. However, it is to be noted that in the regime of weak magnetic fields, rotation can induce stronger effects on the structure of white dwarfs than magnetic fields. In this case, Eq.~\ref{gw} may not be valid.
 

\section{Conclusions} 
 In this work, we performed self-consistent and  relativistic numerical calculations of axisymmetric rotating and magnetized white dwarf structure by means of a pseudo-spectral
method, where the standard stress-energy tensor of a perfect fluid and the electromagnetic field were employed. First, we fixed the baryon stellar mass and we computed  the quadrupole moment of the configuration, which was used to estimate the gravitational wave amplitudes of potential sources. In our case, white dwarfs can be rotating and/or endowed with a poloidal magnetic field. 

We showed that the moment of inertia of white dwarfs increase significantly due to magnetic fields and depend strongly on the stellar mass, where less massive WD's have higher moment of inertia, however, since they are less dense, they reach lower magnetic fields values than massive stars. Moreover, we gave bounds for the radius, magnetic fields, moment of inertia and quadrupole moment of the pulsar white dwarf AR Sco. These results relied on the assumption that the observed luminosity corresponds exactly to the spin-down power undergone by the white dwarf. Although we have assumed this approximation  in
this work, we do not expect qualitative changes in our conclusions, since according to Fig.~\ref{rc_luminosity} a higher luminosity implies just a smaller possible mass for AR Sco.

More importantly, we also found that magnetic white dwarfs might lead to a detectable signal by the DECIGO and BBO gravitational wave detectors. The DECIGO and the BBO, both having the same sensitivity frequency band, perform better than LISA in detecting GW from magnetized  white dwarfs. In addition, we saw that BBO is able to detect GW even from a purely rotating white dwarf (without magnetic fields) with $M_{B}=0.50\,M_{\odot}$ rotating at a frequency of 0.01 Hz, while DECIGO can potentially measure gravitation radiation from the magnetic counterpart of this star. 

A key fact is that the magnetic field (for the star $M_{B}=0.50\,M_{\odot}$) that corresponds to the minimum GW amplitude within the detectable range of DECIGO is close to observed ones, what indicates that magnetized white dwarfs are also likely sources of GW and could be detected by future space-based gravitational wave detectors. Furthermore, based on the GW amplitude, magnetic white dwarfs can emit gravitational radiation so intense as in strongly magnetized neutron stars. 

It is worth mentioning that the detection of white dwarfs with higher frequencies ($\sim$ 1 Hz) would make them  better candidates of gravitational wave sources, since they would lie in the optimal band of DECIDO and BBO interferometers. Note that, we did not run out of all possible combinations of masses, rotation and magnetic field configurations. However, non-magnetized white dwarfs can reach (Keplerian limit) frequencies at most $\sim$ 1.50 Hz, see e.g. \cite{Franzon:2015gda}. As a consequence, we have a natural limit when looking for white dwarfs through gravitational waves emission.

It is well known that simple magnetic field configurations with purely poloidal or purely toroidal components  are unlikely to be stable \citep{tayler1973adiabatic, markey1973adiabatic, flowers1977evolution, braithwaite2006stable}. Note that the quadrupole distortion induced by toroidal magnetic fields would contribute negatively to the total stellar deformation. This is the case because poloidal magnetic fields make the stars more oblate, while toroidal fields make them more prolate. In this context, an estimate of the stellar quadrupole distortion assuming both poloidal and toroidal magnetic field components was perfomed by \cite{wentzel1960hydromagnetic, ostriker1969nature}. Accordingly, the stellar deformation could be reduced by $50\%$ when  a mixture of poloidal and toroidal fields of similar strength are presented in the star. The distortion of the star scale with its quadrupole moment as $\epsilon \sim Q$ \citep{frieben2012equilibrium}. Therefore, we expect that the gravitational wave amplitude, which scale with $Q$ as $h_0\sim Q$, should be only roughly reduced by a factor of 2 compared to those obtained in this work. In addition, magnetized stars seem to carry an external dipole magnetic field as the one modeled in this work. Furthermore, even assuming for simplicity purely poloidal fields, we can have a fair idea of the maximum magnetic field strength that can be reached inside these stars and also understand the effects of strong magnetic fields both on the GW emission and on global structure of white dwarfs.

In the future, we are going to include effects due to microphysics in our calculations as, for example, a more realistic equation of state which includes electron-ion interactions. Moreover, although we do not expect that our findings are going to change qualitatively, different particle composition of the star might lead to different gravitational wave amplitudes.


\section{Acknowledgements}

B. Franzon acknowledges support from CNPq/Brazil, DAAD and HGS-HIRe for FAIR. S. Schramm acknowledges support from the HIC for FAIR LOEWE program. The authors wish to acknowledge the "NewCompStar" COST Action MP1304.

\bibliographystyle{mnras}
\bibliography{biblio}

\end{document}